\documentstyle[aps,prd]{revtex}

\begin{document}

\draft

\title{Non stationary Einstein-Maxwell fields interacting with a superconducting cosmic string}
\author {Reinaldo J. Gleiser\thanks{Electronic address: gleiser@fis.uncor.edu} and Manuel H. Tiglio\thanks{Electronic address: tiglio@fis.uncor.edu} }
\address{\it Facultad de Matem\'{a}tica, Astronom\'{\i}a y F\'{\i}sica,
 Universidad Nacional de C\'{o}rdoba \\ Ciudad Universitaria \\
5000 C\'{o}rdoba, Argentina. }
\maketitle

\begin{abstract}

Non stationary cylindrically symmetric exact solutions of the Einstein-Maxwell equations are derived as single soliton perturbations of a Levi--Civita metric, by an application of Alekseev inverse scattering method. We show that the metric derived by L. Witten, interpreted as describing the electrogravitational field of a straight, stationary, conducting wire may be recovered in the limit of a `wide' soliton. This leads to the possibility of interpreting the solitonic solutions as representing a non stationary electrogravitational field exterior to, and interacting with, a thin, straight, superconducting
cosmic string. We give a detailed discussion of the restrictions that arise when appropriate energy and regularity conditions are imposed on the matter and fields comprising the string, considered as `source', the most important being that this `source' must necessarily have a non-vanishing minimum radius. We show that as a consequence, it is not possible, except in the stationary case, to assign uniquely a current to the source from a knowledge of the electrogravitational fields outside the source. A discussion of the asymptotic properties of the metrics, the physical meaning of their curvature singularities, as well as that of some of the metric parameters, is also included.  
\end{abstract}


\section{Introduction}

The possibility that the Universe has undergone one, or perhaps more, phase transitions through its evolution is an intriguing conjecture with far reaching consequences that has attracted much attention and research in several fields of theoretical physics and astrophysics. An immediate result of imposing the condition of causality on the phase transition is the appearance of topological structures, some of which may be stable, precisely as a consequence of the geometrical properties and nature of the fields involved. In particular, the existence of vortex-like solutions among these stable structures which, in an astrophysical context, may appear as either closed loops or open structures, has given support to the concept of {\em cosmic strings} as cosmological objects possibly playing a fundamental role in the formation of structures in the early Universe. 

This role was further enhanced by E. Witten's proposition that cosmic strings might carry superconducting currents \cite{ewit}. In this context, an interesting development has been the introduction of {\em vortons}, as stable closed loops of superconducting strings (for a recent review see, e.g.,\cite{car}).

In the case of open strings, an extreme  idealization is to consider solutions with perfect cylindrical
symmetry, and even further, that the whole of the string internal structure is restricted to a straight line, which coincides with the axis of cylindrical symmetry of the system. This idealization becomes useful in situations where one is interested in the effect of the string on their surrounding medium, including their associated gravitational and, for superconducting strings,  electromagnetic fields. The theoretical framework in this case would be given by the Einstein-Maxwell equations. Perhaps the simplest solutions of this type, corresponding to a stationary current carrying string, was given by L. Witten, together with a prescription for the interpretation of the string as a source endowed with electric and magnetic currents \cite{lwit}.  

The superconducting currents found by E. Witten arise in the context of an analysis of the solutions of the field equations, and they are not associated with topological constraints. Therefore, for particular models, stationary solutions containing superconducting currents may exist only in a limited range of parameters \cite{par}, or even not at all. On the other hand, if we drop the condition of stationarity, it is clear that these currents may still be excited 
by the presence of electromagnetic fields, just as in the case of a normal superconductor. The interplay between these induced superconducting currents and the surrounding electromagnetic and gravitational fields may be quite complicated, because of the inherent nonlinear nature of the equations that govern their evolution. In particular, one may ask questions that concern the way the current rises and eventually falls when a pulse of electromagnetic radiation is incident upon the string or when the nonconducting state is unstable and currents may be generated spontaneously \cite{pet}. It may then be of interest to find exact solutions, even in the idealized case of cylindrical symmetry, where these questions are considered explicitly.

 In this paper we describe a new family of exact cylindrically symmetric non stationary solutions of the Einstein-Maxwell equations, derived as single soliton perturbations of a Levi--Civita metric \cite{lev}, by an application of Alekseev inverse scattering method \cite{ale}, and analyze to what extent they may be interpreted as corresponding to the electrogravitational field exterior to a superconducting cosmic string. The paper is organized as follows. 
In Section II we derive the solution of Alekseev equations. In Section III we present explicit expressions for the metric and electromagnetic potential. An analysis of the spacetimes described by our solutions is given in Section IV, and complemented in several Appendices.  In Section V we show that we can recover the stationary axially symmetric electrovacuum metric as a singular limit of the solitonic solutions, leading to the idea that some of the general, time dependent solutions, can be interpreted, for some interval of time, as describing a region of spacetime near the core of a superconducting cosmic string interacting with Einstein Maxwell fields. This possibility is analyzed in detail in Section VI, where we discuss, without reference to a particular model, certain restrictions on the metric parameters that arise when we consider the metric as describing the spacetime in the region exterior to a cylindrically symmetric source, and impose certain energy and regularity conditions on the stress-energy-momentum of the matter and fields contained in the source. In particular, in the time dependent case, and in contrast to what happens in the stationary case, we find that it is not possible to assign a unique electric current to our electrovacuum metrics, a result of obvious relevance to the physical interpretation of the solutions. The consequences of the results obtained in Section VI are further discussed in Section VII, which contains also some final comments. We also include three appendices. Appendix A gives results for the asymptotic behaviour at timelike infinity. The nature of the singularities and some possible physical and geometrical interpretations thereof are discussed in Appendix B. Finally, the vacuum and diagonal subfamilies of metrics are described in Appendix C.

\section{The Alekseev equations for the seed metric}

The family of exact solutions of the Einstein-Maxwell equations considered in this paper were obtained by an application of Alekseev's inverse scattering method to a suitable `seed' metric. A complete description of this method can be found in \cite{ale} and a review on solitonic solutions in \cite{ver}. The spacetime external to a stationary superconducting string can be approximated by that of conducting wire and this in turn reduces to the Levi-Civita spacetime when the current is set to zero (see the discussion in Section VI). With these facts in mind we chose the Levi-Civita spacetime as seed, written in the form
\begin{equation}
ds^2 = C \rho^{(\Delta^2-1)/2}\left(dt ^2 - d\rho ^2\right) - \rho ^{1+\Delta} dx_2^2 - \rho ^{1-\Delta} dx_3^2 \; ,
\label{l-c}
\end{equation}
where $C$ and $\Delta$ are constants. Once the seed is chosen, the Alekseev method requires that we find a $3 \times 3$ complex matrix $\Psi$, which is a solution of the equations
$$ 
\partial_{\mu} \Psi = \Lambda_{\mu}^{\;\;\nu } U_\nu \Psi  \; ,
$$
where $(\mu, \nu = \rho,t)$, $U_{t}$ and $U_{\rho}$ are $3 \times 3$ complex matrices, determined by the seed metric, and $\Lambda_{\mu}^{\;\; \nu}$ some functions of $t$ and $\rho$. For the particular form of (\ref{l-c}), (see \cite{ale} for details) we obtain
$$
\label{Psi1}
\partial_{\rho} \Psi =  \frac{1}{2i\left[ \left( i\omega+t \right)^2 - \rho^2 \right] }
\left( \begin{array}{ccc}
-i\rho(\Delta-1) & \left( i\omega+t\right) (\Delta+1) \rho^\Delta & 0 \\ 
\left( i\omega+t\right) (\Delta-1) \rho^{-\Delta } & i\rho(\Delta+1) & 0 \\ 
0 & 0 & 0
\end{array} \right) \Psi
$$
and
$$
\label{Psi0}
\partial _t\Psi =  \frac{1}{2i\left[ \left( i\omega+t\right)^2 - \rho^2 \right] }
\left( \begin{array}{ccc}
i\left( i\omega+t\right) (\Delta -1) & -\rho^{\Delta+1}(\Delta+1) & 0 \\ 
-(\Delta-1) \rho^{-\Delta+1} & -i\left( i\omega+t\right) (\Delta+1) & 0 \\ 
0 & 0 & 0
\end{array} \right) \Psi \; .
$$

The general solution of these equations, satisfying certain conditions required by the method, may be written in the form
$$
\Psi (\rho,t,\omega )=
\left( \begin{array}{ccc}
k_1\Phi _{22}+k_2\Phi _{23} & k_3\Phi _{22}+k_4\Phi _{23} & 0 \\ 
k_1\Phi _{32}+k_2\Phi _{33} & k_3\Phi _{32}+k_4\Phi _{33} & 0 \\ 
0 & 0 & 1
\end{array}\right) \; ,
$$
with $k_i \;\; (i=1 \ldots 4)$ arbitrary complex constants, and
$$
\Phi _{22}(\omega )=i\frac{\rho^{\Delta /2}}{\sqrt{2}}\left( \frac{\sinh (\Delta
\ln \Lambda )}{\sigma _{+}}+\frac{\cosh (\Delta \ln \Lambda )}{\sigma _{-}} \right) \; ,
$$

$$
\Phi _{23}(\omega )=\frac{\rho^{\Delta /2}}{\sqrt{2}}\left( \frac{\cosh (\Delta \ln
\Lambda )}{\sigma _{+}}+\frac{\sinh (\Delta \ln \Lambda )}{\sigma _{-}} \right)  \; ,
$$

$$
\Phi _{32}(\omega )=\frac{\rho^{-\Delta /2}}{\sqrt{2}}\left(\frac{\sinh (\Delta
\ln \Lambda )}{\sigma _{+}}-\frac{\cosh (\Delta \ln \Lambda )}{\sigma _{-}} \right)  \; ,
$$

$$
\Phi _{33}(\omega )=-i\frac{\rho^{-\Delta /2}}{\sqrt{2}}\left(\frac{\cosh (\Delta
\ln \Lambda )}{\sigma _{+}}-\frac{\sinh (\Delta \ln \Lambda )}{\sigma _{-}} \right)  \; ,
$$
where
$$
\sigma_{+} = \sqrt{i\omega+t+\rho} \;\;\; , \;\;\;\sigma _{-}=\sqrt{i\omega+t-\rho}\;\;\; , \;\;\;\; \Lambda =\left( \frac{\sigma_{-}+\sigma _{+}}{\sigma _{-}-\sigma _{+}} \right)^{1/2} \; .
$$

If we restrict to single soliton transformations, we may choose $\omega$ real without loss of generality, as in this case the imaginary part of $\omega$ may be absorbed in a redefinition of the origin of $t$. We also choose $\omega >0$, and, for the multivalued functions that arise in the new metric, we choose the branch cut  along the negative real axis in the corresponding complex plane. Given these conventions, we define the real quantities  $\sigma >0$ and $\phi$ by
\begin{equation}
\sigma_+ + \sigma_- = \sqrt{2\rho}\sigma^{1/4}e^{i\phi/2}  \; . \label{funciones}
\end{equation}

\section{The metric and the electromagnetic potential}

Once $\Psi(\rho,t,\omega)$ is found, the remaining steps in the construction of the metric are purely algebraic, and may be straightforwardly handled using computer algebra. If we write the final metric in the form
$$
ds^2=f(dt^2 - d\rho^2) + g_{22}dx_2^2 + g_{33}dx_3^2 + 2g_{23}dx_2dx_3 \; ,
$$
the coefficients are given by
$$
f = \frac{C'}{{\cal H}}\rho^{(\Delta^2-1)/2}\sigma {|\cal{D}|}^2 \; ,
$$
\begin{eqnarray*}
g_{22} & = & - \frac{\rho^{1+\Delta}}{{|{\cal D}|}^2} \left[\left(\sqrt{\sigma}-\frac{1}{\sqrt{\sigma}}\right)^2 \left( \sin{(\phi+\delta)}\left(\sin{(\phi+\delta)}-\frac{q}{L_0}\right) + \frac{q^2}{4L_0^2}\right)  \right. \nonumber \\ 
 & & + \left.\sin{\phi}^2 \left(   \left(\frac{L_0}{\sigma^{(\Delta+1)/2}}+\frac{\sigma^{(\Delta+1)/2}}{L_0}\right)\left(\frac{L_0}{\sigma^{(\Delta+1)/2}}+\frac{\sigma^{(\Delta+1)/2}}{L_0}+\frac{2q}{L_0}\right) + \frac{q^2}{L_0^2}\right) \right]  \; ,
\end{eqnarray*} 
\begin{eqnarray*}
g_{23} & = & -\frac{2\omega}{{|{\cal D}|}^2 \sqrt{\sigma}} \left[ L_0\left( \frac{1}{\sigma^{\Delta/2}} \left(\sin{(\phi-\delta)}+\sigma \sin{(\phi+\delta)}\right) \right) + \frac{2q}{L_0} \sqrt{\sigma} \sin{\phi} \cos{\delta}  \right. \nonumber \\
& &   \left. + \frac{\sigma^{\Delta/2}}{L_0} \left(\sin{(\phi+\delta)} + \sigma \sin{(\phi-\delta)} \right) + \frac{q}{2L_0} \left( 1-\sigma \right) \left( \frac{L_0}{\sigma^{\Delta/2}} -\frac{\sigma^{\Delta/2}}{L_0} \right) \right] \; ,
\end{eqnarray*} 
\begin{eqnarray*}
g_{33} & = & -\frac{\rho^{1-\Delta}}{{|{\cal D}|^2}} \left[\left(\sqrt{\sigma}-\frac{1}{\sqrt{\sigma}}\right)^2 \left( \sin{(\phi-\delta)}\left(\sin{(\phi-\delta)}+\frac{q}{L_0}\right) + \frac{q^2}{4L_0^2}\right) \right. \nonumber \\ 
& & \left. + \sin{\phi}^2 \left(   \left(\frac{L_0}{\sigma^{(\Delta-1)/2}}+\frac{\sigma^{(\Delta-1)/2}}{L_0}\right)\left(\frac{L_0}{\sigma^{(\Delta-1)/2}}+\frac{\sigma^{(\Delta-1)/2}}{L_0}+\frac{2q}{L_0}\right) + \frac{q^2}{L_0^2} \right) \right] \; ,
\end{eqnarray*}
where
$$
\delta = \Delta \phi + \delta_0 \; ,
$$
$$
{\cal H} = (1-\sigma)^2 + \frac{16\omega^2\sigma^2}{(1-\sigma)^2\rho^2} \; ,
$$
$$
{\cal D} = \left(\sqrt{\sigma}-\frac{1}{\sqrt{\sigma}}\right)\left(\sin{\delta}-\frac{q}{2L_0}\cos{\phi}\right) + i \left[ \frac{L_0}{\sigma^{\Delta/2}}+\frac{\sigma^{\Delta/2}}{L_0}+\frac{q}{2L_0}\left(\sqrt{\sigma}+\frac{1}{\sqrt{\sigma}}\right)\right] \sin{\phi} \; ,
$$
and $\delta_0 \;,\; q \;,\; L_0$ and $C'$ are arbitrary real constants, which are positive, except for $\delta_0$, which can also take negative real values.
 
 The complex potential for the selfdual electromagnetic tensor is given by
 \begin{eqnarray*}
\Phi_2 & = & \frac{\omega\sqrt{2q}}{\cal D} \rho^{(\Delta-1)/2}e^{i(4\theta_d+2\delta_0+\pi)/4} \left[ \cos{\left(\frac{\phi+\delta+\pi/2}{2}\right)} \left( -\frac{\sigma^{(\Delta+1)/4}}{L_0}+\frac{1}{\sigma^{(\Delta+1)/4}} \right)  \right. \nonumber \\
& &  \left. + i \sin{\left(\frac{\phi+\delta+\pi/2}{2}\right)}\left( \frac{\sigma^{(\Delta+1)/4}}{L_0} + \frac{1}{\sigma^{(\Delta+1)/4}}   \right)    \right] \; ,
\end{eqnarray*}
\begin{eqnarray*}
\Phi_3 & = & \frac{\omega\sqrt{2q}}{\cal D} \rho^{(-\Delta-1)/2}e^{i(4\theta_d+2\delta_0-\pi)/4} \left[ \cos{\left(\frac{-\phi+\delta-\pi/2}{2}\right)} \left( -\frac{\sigma^{(\Delta-1)/4}}{L_0}+\frac{1}{\sigma^{(\Delta-1)/4}} \right)  \right. \nonumber \\
& &  \left. - i \sin{\left(\frac{-\phi+\delta-\pi/2}{2}\right)}\left( \frac{\sigma^{(\Delta-1)/4}}{L_0} + \frac{1}{\sigma^{(\Delta-1)/4}}   \right)    \right] \; .
\end{eqnarray*}
 
We notice that $q$ provides the scale for the electromagnetic tensor,  $\theta_d$ defines dual rotations of the electromagnetic field, and  $L_0$ may be considered as the `polarization parameter' for the vacuum gravitational field. Solutions with $\Delta$ and $-\Delta$ are locally isometric, as long as $L_0 \neq 0$. This can be seen performing the interchanges $L_0 \leftrightarrow 1/L_0$ and $x_2 \leftrightarrow x_3$. 
 
The electromagnetic field vanishes for any value of $\Delta$ if $q=0$, and  the solutions correspond to vacuum. In these cases the metrics reduce to those obtained as a solitonic perturbation of the Kasner metric, using the Belinski-Zakharov inverse scattering method and two complex poles \cite{ver,ver2}. This can be checked by simply noticing that the functions $\sigma$ and $\phi$ as defined in Eq.(\ref{funciones}) coincide with the ones defined in \cite{ver2},  and that by substituting $q= 0$ in our solution one obtains the metric in the same explicit form as that given in that reference.

In the next Section we study some general limits of the metrics, for different ranges of the parameters, and the behaviour near the `axis' $\rho=0$ , at `spacelike infinity' $\rho \rightarrow \infty$ , on the future `null cones', and for large $|t|$. Some detailed expressions are given in the Appendices. In Section V we consider the stationary limits of the metric. These will be important in the discussion of its physical interpretation given in the following Sections. Readers interested only in the physical interpretation of the metrics and their relation to superconducting cosmic strings may skip the next section. 

\section{Analysis of the spacetime}

In this section we analyze the behaviour of the metric near the symmetry axis, at spatial, timelike and null infinity, assuming $q \neq 2L_0 \sin{\delta _0}$. As can be seen in Appendix B, if this condition is not satisfied the behavior can be quite different.

At spatial infinity, i.e., fixed $t$ and $\rho \rightarrow \infty$, the metrics approach the seed for all values of $\Delta$. Namely,
$$
ds^2 \approx \frac{C'(1+q+L_0^2)^2}{4}\rho^{(\Delta^2-1)/2} (dt^2-d \rho ^2) - \rho^{1+\Delta} dx_2 ^2 - \rho^{1-\Delta} dx_3  ^2 \; . 
$$

The metrics also approach the seed in the asymptotic light cone, for all values of $\Delta$. Specifically, defining $u=\rho+t$ and $v=\rho-t$, for fixed $v$ and $u \rightarrow \infty$ we obtain:
\begin{eqnarray}
ds^2 & \approx & \frac{C' (u/2)^{(\Delta^2-1)/2}}{8L_0^2 \left[ v ^2 + \omega ^2 + v  (v ^2 + \omega ^2)^{1/2} \right] } \left\{ 2 \left( q - 2L_0 \sin{\delta_0} \right)^2 \left[ v ^2 + \omega ^2 + v  (v ^2 + \omega ^2)^{1/2} \right] + \right.  \label{cone} \\
& & \left. \omega ^2 \left[ \left( L_0 ^2+ 1+ q \right)^2 - \left( q - 2L_0 \sin{\delta_0} \right)^2  \right] \right\} dudv - \left( \frac{u}{2} \right)^{1+\Delta} dx_2^2 - \left( \frac{u}{2} \right)^{1- \Delta} dx_3^2 \; , \nonumber
\end{eqnarray}
which can be rewritten as the seed metric by a simple change of variable on $v$.

Contrary to the previous limits, at timelike infinity, i.e., fixed $\rho$ and $t \rightarrow \infty$, the behavior depends strongly on the value of $\Delta$. Details are given in Appendix A. It is found that the metrics approach the background only for $0<\Delta < 1$, while a singular behavior results for other values of $\Delta$. 
 
Regarding the behavior of the metric near the symmetry axis, i.e., for $\rho \simeq 0$, we also find a qualitative dependence on $\Delta$. It can be seen that for $\Delta >2$ we have

\begin{equation}
ds ^2 \approx  \frac{C' \rho ^{(\Delta ^2-1)/2} \sigma ^{\Delta -1} \sin^2{\phi } }{L_0 ^2} (dt^2 - d\rho ^2) -  \sigma \rho^{\Delta +1} dx_2 ^2 - \frac{\rho ^{1- \Delta}}{\sigma } dx_3 ^2   \label{axis1}
\end{equation}
with
\begin{equation}
\sigma \approx \frac{4(\omega ^2 + t^2)}{\rho ^2} \;\;\; , \;\;\; \sin^2{\phi } \approx \frac{\omega ^2}{t^2 + \omega ^2}  \; ,\label{sigma1}
\end{equation}
and, interestingly, as we shall see later, if one does not approximate $\sigma$ and $\phi$ by their values near $\rho =0$, i.e., (\ref{sigma1}), then (\ref{axis1}) is an exact vacuum diagonal metric {\em for all spacetime}. In other words, when $\Delta>2$ the metric near the axis behaves as in the vacuum, diagonal (solitonic) case.

Concerning the regularity or quasiregularity of the symmetry axis, these conditions can be attained only if $\Delta =1$ or $\Delta =3$ . It is interesting to note that these two cases are precisely those in which there is boost invariance along the $z$ direction at spatial infinity. We proceed to discuss these two cases.

When $\Delta = 1$, and $q \neq 2L_0 \sin{\delta _0}$, our solutions coincide with those obtained in \cite{dgn90}. Although this is not immediate from the expressions for the metric and electromagnetic fields, it follows essentially by construction, since in that reference Minkowski spacetime was used as seed ($\Delta=1$). One can also see that the condition $q \neq 2L_0 \sin{\delta _0}$ is $\xi ^2 =1$ in the notation of \cite{dgn90}. We will not give the explicit relations between the constants and functions used in this paper and the ones used in \cite{dgn90}, but just summarize some results. Performing the change of variables:

\begin{equation}
\theta = x_2 \;\;\; , \;\;\; z = -4 \omega \left( q - 2L_0 \sin{\delta _0} \right)^{-1}  x_2 + x_3 \; , \label{new}
\end{equation}
(which is valid as long as $q \neq 2 L_0 \sin{\delta _0}$) the metric tends to the seed (flat spacetime) at spatial and timelike infinity. At future null infinity the metric also tends to flat spacetime, and there is an outgoing flux of C - energy (this flux can be straightforwardly calculated from (\ref{cone}) with $\Delta =1$). The metric can be chosen regular or quasiregular near the axis, with a deficit angle that is constant in time. At spacelike infinity the deficit angles is $< 2\pi$ (indeed, it cannot exceed this value if the spatial sections are to be noncompact and the intrinsic metric geodesically complete, as is be discussed in Appendix B). Thus, this solution represents a   non supermassive gauge cosmic string interacting with Einstein - Maxwell fields, with the property that the deficit angle at the axis is constant in time, and thus, when de interaction ceases (at future timelike infinity) the string has the same mass per unit of length as it had before the interaction (past timelike infinity).

To check that the axis can be made either regular or quasiregular when $\Delta =3$ , it suffices to substitute this value in (\ref{axis1}). Then one obtains:
$$
ds^2 = \frac{1}{L_0^2} C' \omega ^2 16 (t^2 + \omega ^2)  \left( -dt^2 +d\rho ^2 - \rho ^2 \frac{L_0^2}{4 C' \omega ^2} dx_2 ^2 \right) - \frac{1}{4(t^2 + \omega^2)} dx_3 ^2 \; ,
$$
which proves the previous assertion. The deficit angle is given by
$$
\delta \phi = 2\pi \left(1 - \frac{L_0}{2 C'^{1/2} \omega} \right) \; ,
$$
and the same feature we have already mentioned for $\Delta =1$ is present: the deficit angle is independent of time.

Returning to the general case (arbitrary $\Delta$), we have already indicated that at future timelike infinity the metric approaches the seed iff $0\leq \Delta<1$.  In these cases, however, as we will show in Section VI, the seed that is approached by the metric corresponds to a source that violates the Strong Energy Condition,  and thus these cases require unphysical source and are probably of no physical interest. Finally, for $\Delta >1$ the norm of one of the killing vector diverges when $t \rightarrow \infty$ , although we notice that this singularity is not reached by  observers with constant $\rho$ , $\phi$ and $z$, since their proper time approaches infinity when $t \rightarrow \infty$ .

We remark that, on account of Eqs.(\ref{latea}-\ref{lated}) of Appendix A, some of the asymptotic expressions at future timelike infinity are not valid when $q = 2 L_0 \sin{\delta _0}$, and neither is the change of variables in the $\Delta =1$ case that led to a quasiregular axis (Eq. (\ref{new})). This point is considered separately in Appendix B. As we shall see, the analysis of these cases will help in understanding the nature of the singularities that develop at late times.

\section{The stationary limit}

From  general properties of the inverse scattering method, we know that we essentially recover the background metric in the limit $\omega \rightarrow 0$. In our case this is a stationary vacuum metric. On the other hand, if we consider the soliton metric, we notice that $t$ appears only in expressions of the form $i \omega \pm t$. This implies that we may expect the metric to be approximately stationary (in the sense that it depends only weakly on $t$) in the region $|\omega| \gg |t|$ , and this may hold either because $|t|$ is  small, or $|\omega|$ is large, or both. Moreover, $\omega$ may be related to the `width' of the soliton, so that large $\omega$ corresponds to a wide, and therefore slowly varying, soliton. This suggests that we analyze the limit $\omega \rightarrow \infty$, and, as we show below, it turns out that it is indeed possible to recover the stationary electrovacuum solution, (\ref{lwitten}), as a singular limit when $\omega \rightarrow \infty$ .  

One way to obtain the stationary electrovacuum solution is to first take the limit $L_0 \rightarrow 0$. Then we define $j$ by $q=2^{\Delta} \omega ^{\Delta -1}/ j^2$, and perform the following change of coordinates: 
$$
x_3 \rightarrow \frac{2^{2 \Delta} \omega^{2\Delta -2}}{q^2 j^4}(x_3 + 2 \omega j^2 x_2) \;\;\; , \;\;\; x_2 \rightarrow \frac{q^2 j^4}{2^{2 \Delta} \omega^{2\Delta -2}} x_2 \; .
$$
Then, for $\omega \rightarrow \infty$ (and $C' (2 \omega)^{2\Delta -2}/j^4 \rightarrow C''$ with $C''$ finite) one explicitly obtains

\begin{eqnarray}
ds^2 &\approx & C'' \rho ^{(\Delta ^2-1)/2} \Gamma^2 (dt^2-d \rho ^2) - \rho^{1+\Delta}\Gamma^2 d\phi^2 - \rho^{1-\Delta}\Gamma^{-2} dz^2 \nonumber \\ & & + \mbox{ time dependent terms of order }  {\omega}^{-1} \; ,\nonumber 
\end{eqnarray}
where $ \Gamma= \left(1+ j^2 / \rho ^{\Delta -1} \right)$. Thus, the metric approaches (\ref{lwitten}) in the limit $\omega \rightarrow \infty$ .

A simple interpretation of this result is that, at least for a certain range of parameters, and for sufficiently large $\omega$, in the region $\rho \ll |\omega|$, $|t|\ll |\omega|$, i.e., close to the symmetry axis and near $t=0$, the metric describes approximately a superconducting string with a slowly varying current. On this account, it would be natural to interpret the full metric as describing the spacetime outside a superconducting string, interacting with a time dependent electromagnetic field, and, therefore, carrying a time dependent current. A closer analysis shows, however, that there are some subtle issues that arise when we require that this exterior metric be matched to a `source' satisfying some physically acceptable conditions. This analysis is carried out in the following section.

\section{The relation between the exterior metric and the internal structure of cylindrical sources}

As we saw in Section IV,  in general, the metrics obtained in the Section III are singular for $\rho \rightarrow 0$. This singularity corresponds to the fact that certain curvature scalars are unbounded as we approach $\rho=0$ . This implies that the metrics cannot be extended to include the symmetry axis $\rho=0$ . On the other hand, it is legitimate to ask if these metrics may be considered as the electrogravitational field external to some source, and to require that the source be regular, with a well defined axis of symmetry.  Discussions of this type of problem, for different types of cylindrically symmetric sources, but restricting to vacuum exterior metrics, can be found, e.g., in \cite{cil}, and in the analysis of superconducting gauge strings in \cite{mos}. Here we consider again the problem, reviewing some well known results, and adding others which are of interest for the present analysis. 

Electrovacuum metrics with `full cylinder symmetry' \cite{tho}, exterior to an infinite stationary cylinder, have been obtained by several of authors \cite{tho,dem,aut}. An interesting approach using the Rainich conditions can be found in \cite{lwit}. The general form of these metrics can be given as
\begin{equation}
ds^2 = C \rho ^{(\Delta ^2-1)/2} \Gamma^2 (dt^2-d \rho ^2) - \rho^{1+\Delta}\Gamma^2 d\phi^2 - \rho^{1-\Delta}\Gamma^{-2} dz^2  \; , \label{lwitten}
\end{equation}
where $\Gamma = \left(1+j^2/\rho^{\Delta-1}\right)$, and the topology is defined by choosing the coordinates as the usual cylindrical ones, i.e., $t \in {\cal R}$ , $\rho >0$ , $z \in {\cal R}$ and $\phi \in [0 , 2\pi ]$. $\Delta$ , $C$ and $j$ are three arbitrary constants, related to the existence of three independent curvature scalars\cite{lwit}. For the discussion that follows it is convenient to write this metric in the form
\begin{equation}
ds^2 = \rho ^{(\Delta ^2-1)/2} \Gamma^2 (dt^2-d \rho ^2) - \rho^{1+\Delta}\Gamma^2 \left( 1 - \frac{\delta}{2\pi} \right)^2 d\phi^2 - \rho^{1-\Delta}\Gamma^{-2} dz^2   \label{lwittenb}
\end{equation}
with $0 \leq \delta <2\pi$ .

We consider first the vacuum case, i.e., $j=0$. Then the metric reduces to that of Levi-Civita. Solutions that differ only in the sign of $\Delta$ are locally isometric. There is boost invariance along the $z$ direction if $\Delta =1$ or $\Delta=3$ . If $\Delta \neq 1$ there are curvature scalars that diverge for $\rho \rightarrow 0$ and for $\rho \rightarrow \infty$ . For $\Delta > 0$, the singularity for $\rho \rightarrow 0$ may be reached by causal geodesics with finite affine parameter, while this is not possible for $\rho \rightarrow \infty$. For $\Delta =1$ the metric is (locally) flat and regular except for a conical singularity on the symmetry axis if $\delta \neq 0$ . Particles and photons are repelled by the singularity if $\Delta < 1$ .

In the electrovacuum case ($j \neq 0$), when $\Delta \neq -1$ we also find diverging curvature scalars when $\rho \rightarrow 0$ and $\rho \rightarrow \infty$ and, just as in the vacuum case, the singularity for $\rho \rightarrow 0$ ($\rho \rightarrow \infty$) is reachable (unreachable) by causal geodesics with finite affine parameter. There is certain relationship between these singularities and the presence of an electric current (proportional to $j$) confined to the symmetry axis (see \cite{lwit}). When $\Delta=-1$ the metric can be made regular everywhere, (including the symmetry axis) and describes a spacetime filled with a magnetic field, the Melvin magnetic universe (the coordinates $z$ and $\phi$ must be interchanged for this solution) \cite{mel}. None of these solutions is boost invariant along the $z$ axis. If $1< \Delta < 3$ particles and photons are repelled (see  \cite{dem}) for
$$
\rho<\left[ j^2\left( \frac{3-\Delta }{1+\Delta }\right) \right] ^{1-\Delta} \; .
$$

On physical grounds, we expect the singularity on the symmetry axis to be largely related to the extreme idealization of a `source' (be it matter, or electric current, or both), being confined to a cylinder of vanishing thickness.  It is then of interest to inquire what are the conditions that arise when we want to match the metric corresponding to a cylinder of matter of finite radius ${\cal R}_0$ with an exterior vacuum or electrovacuum metric, in such a way that the metric is regular inside the cylinder. The purpose of this exercise is twofold: first, to relate the parameters characterizing the external metric to the type of matter contained in the cylinder, and, second, to show that, as might be expected, the cases where particles and photons are repelled by the cylinder correspond to matter violating some energy condition. 

We assume that the interior metric is diagonal, stationary, axially symmetric and everywhere regular (in particular, on the symmetry axis). In this case, using `standard coordinates' \cite{tho}, it may written in the form 
\begin{equation}
ds^2=e^{2(\gamma -\psi )}\left( dt^2-d\rho^2\right) -e^{2\psi}dz^2-\alpha ^2e^{-2\psi }d\phi^2  \label{standard}
\end{equation}
where the ranges for the coordinates are the same as for (\ref{lwitten}). The axis is regular if, for $\rho \approx 0$,
\begin{equation}
ds^2 \approx (dt^2 - d\rho ^2) - dz^2 - \rho ^2 d\phi ^2  \; .   \label{regularity}
\end{equation}

From Einstein's equations for the metric (\ref{standard}) we obtain the following relations:
 
\begin{equation}
\label{relat1}
\alpha ^{\prime \prime }= - {\cal T}_1 \;\;\;\;,\;\;\;\;
 \left( \alpha (\gamma -\psi)^{\prime }\right) ^{\prime } = {\cal T}_2
\;\;\;\;,\;\;\;\;
 \left( \alpha (\gamma - 2\psi)^{\prime }\right) ^{\prime } = {\cal T}_3 \; ,
\end{equation}
where

$$
{\cal T}_1 = 8\pi G\left( \tilde{{\cal T}}_t^{\;\;t}+\tilde{{\cal T}} _\rho^{\;\;\rho}\right)  \;\;\; , \;\;\; {\cal T}_2 = 8\pi G\left( \tilde{{\cal T}}_t^{\;\;t}-\frac 12\tilde{{\cal T}}\right) \;\;\; , \;\;\; {\cal T}_3 =  8\pi G\left( \tilde{{\cal T}}_t^{\;\;t}- \tilde{{\cal T}}_z^{\;\;z} \right)  \; ,
$$
with ${\cal T}_i^{\;\;j}$ the energy - momentum tensor, $\tilde{{\cal T}}_i^{\;\;j} \equiv \left| g\right| ^{1/2} {\cal T}_i^{\;\;j}$ , $\tilde{{\cal T}} \equiv \tilde{{\cal T}}_t^{\;\;t}+\tilde{{\cal T}}_\rho ^{\;\;\rho}+\tilde{{\cal T}}_\phi ^{\;\;\phi }+\tilde{{\cal T}}_z^{\;\;z}$, and primes indicate derivatives with respect to $\rho$ . 

If we impose the Darmois matching conditions, i.e., continuity of the first and second fundamental forms on the joining surface at $\rho ={\cal R}_0$, then we must require that $\alpha$, $\gamma$ and $\psi$ be continuous with continuous first derivatives. Integrating the equations (\ref{relat1}) in the plane defined by the coordinates $\{ \rho \;, \; \phi \}$, in a disk of radius ${\cal R}_0$, and using (\ref{lwitten}) and (\ref{regularity}), we find the following conditions for matching the interior metric with (\ref{lwittenb}) at $\rho = {\cal R}_0$,

\begin{equation}
 \int_0^{2\pi}  \int_0^{{\cal R}_0} {\cal T}_1 \;   d \rho \; d \phi  = \delta \; ,   \label{deficit}
\end{equation}

\begin{equation}
\int_0^{2\pi} \int_0^{{\cal R}_0}  {\cal T}_2 \;   d \rho \; d \phi   =  
{ (2\pi - \delta ) (\Delta-1)   [(\Delta+1) {\cal R}_0^{\Delta-1} + (\Delta-3) j^2] \over  4  ({\cal R}_0^{\Delta-1} +j^2)}   \; ,
\label{SEC}
\end{equation}

\begin{equation}
 \int_0^{ 2\pi} \int_0^ {{\cal R}_0} {\cal T}_3 \;    d \rho \; d \phi  =  { (2\pi -\delta ) (\Delta-1)   [(\Delta+3) {\cal R}_0^{\Delta-1} + (\Delta-3) j^2] \over 4  ({\cal R}_0^{\Delta-1} +j^2)}     \; .
\label{cuerda}
\end{equation}

Given the interior metric, these equations may be used to obtain $\Delta$ , $j$ and $\delta$. Let us analyze first the vacuum case. Consider Eq.(\ref{SEC}). If the Strong Energy Condition (SEC) is satisfied, the left hand side of this equation should be non negative. On the other hand, the right hand side is negative definite for $\Delta <1$ , and the SEC must be violated at some set (of finite measure) inside the cylinder. Similarly, from (\ref{cuerda})  we notice that for a flat exterior metric  ($\Delta=1$), the tension along the symmetry axis must equal the linear mass-energy density. For a ${\cal U} (1)$ gauge string coupled to gravity (in the nonsupermassive, stationary and infinitely long case), the condition ${\cal T}_3 =0$ holds at each point of the spacetime, so taking the limit ${\cal R}_0 \rightarrow \infty$ in  (\ref{cuerda}) one obtains that at large distances from the core the spacetime approaches flat spacetime with a deficit angle, that is obtained by taking the same limit in the left hand side of (\ref{deficit}) \cite{gar}. 

We have a similar situation for the electrovacuum case. The SEC must be violated if 
$$
1<\Delta <3 \;\;\;\;\; \mbox{and} \;\;\;\;\; {\cal R}_0< \left[ j^2\left( \frac{3-\Delta }{1+\Delta }\right) \right] ^{1-\Delta} \; ,
$$
which is the condition for `repulsion' of test particles. This means that the radius of the source cannot be made arbitrarily small. This, as we shall show below, has important implications for the interpretation of the physical nature of the source in terms of electric currents. For a ${\cal U} (1) \times {\cal U} (\tilde{1})$ superconducting gauge string the fields far away from the core are purely magnetic and thus the spacetime approaches (\ref{lwittenb}) with $\Delta >1$. Furthermore, in the cases of interest $(\Delta -1) \ll 1$ and thus $\rho ^{\Delta ^2 -1} \approx 1$ even on cosmological scales. This means that the spacetime is approximately flat and $\delta$ approximately measures the angle of light bending by the string \cite{mos,bab}, as in the nonconducting case.

Now consider the Maxwell equations for the self dual electromagnetic field tensor, 
\begin{equation}
d {\cal F}^{\dagger} = 4 \pi j^{\dagger} \; . \label{dual1}
\end{equation}

In general, the complex current 4-vector $j^{\dagger}$ is a linear combination of electric and magnetic parts. This means that in a local Lorentz frame (\ref{dual1}) may be written in the form
\begin{eqnarray}
\label{elemag}
\vec{\nabla} \cdot (\vec{{\cal E}} + i\;\vec{{\cal B}}) & = & 4 \pi (\rho _e + i\; \rho _m ) \; , \nonumber \\
 \partial _t ( \vec{{\cal E}} + i \; \vec{{\cal B}}) + i \;  \vec{\nabla} \times (\vec{{\cal E}} + i \; \vec{{\cal B}}) &  = & - 4 \pi (\vec{j}_e
+ i \; \vec{j}_m ) \; , 
 \end{eqnarray}
where $ i $ is the imaginary unit, all the other quantities are real, and we must have 
\begin{equation}
\label{conserv}
\vec{\nabla} \cdot (\vec{j}_e+ i \;\vec{j}_m ) =  - \partial_t \left(\rho _e  + i \;   \rho _m   \right) 
\end{equation}
as an integrability condition. In these equations $\rho_e$ and $\vec{j}_e$ represent, respectively, the electric charge density and current density, while $\rho_m$ and $\vec{j}_m$ are the corresponding magnetic counterparts. In the case where everywhere we have
$$
\vec{j}_e = \alpha \;\vec{j}_m \;\;\; ,\;\;\; \rho _e  = \alpha \; \rho _m   \; , 
$$
with $\alpha$ some real constant, the magnetic part of $j^{\dagger}$ may be eliminated by a redefinition (`dual rotation') of the fields $\vec{{\cal E}}$ and $\vec{{\cal B}}$. However, if ${\cal F}^{\dagger}$ appears in the context of gauge field theories, we may envisage situations where $j^{\dagger}$ contains a non trivial magnetic part. Since in our particular problem we are only considering the vacuum region, where $j^{\dagger}=0$ , the question that naturally arises is to what extent can we obtain in this case information on the nature of the source $j^{\dagger}$ by considering only the fields in that region. To answer this question we notice that if we apply Stoke's theorem to a $3$ dimensional hypersurface $\Sigma$ with boundary $S$, we have

\begin{equation}
\int_\Sigma j^{\dagger}=\frac 1{4\pi }\int_S{\cal F} ^{\dagger} \; .  
\label{stokes}
\end{equation}

Restricting to cylindrical symmetry, and assuming that there is regular `source region' for $0 \leq \rho < {\cal R}_0$, with ${\cal R}_0$ some fixed `radius', where the current $j^{\dagger}$ is also regular, we choose $\Sigma$ as a $3$ cylinder on a constant $z$ surface, with arbitrary radius ${\cal R} \geq {\cal R}_0$ , and boundary given by  $S=S_1\cup S_2\cup S_3$ , where $S_1$ and $S_2$ are disks of radius ${\cal R}$ , for constant $z$, taken respectively at times $t$ and $t+\tau$, and $S_3$ is a cylindrical 2-surface, at the same constant $z$ , with radius ${\cal R}$, and `height' $\tau$. Then, from (\ref{stokes}), we have
\begin{equation}
\int_\Sigma j^{\dagger} = \frac 1{4\pi } \left( \int_{S_3}{\cal F}^{\dagger}_{02} dt d\phi+\int_{S_2}{\cal F}^{\dagger}_{12} d\rho d\phi - \int_{S_1}{\cal F}^{\dagger}_{12} d\rho d\phi \right)   
\label{corriente}
\end{equation}
The term on the left hand side of (\ref{corriente}) is the total current ${\cal I}$ in the $z$ direction, integrated in time from $t$ to $t +\tau$ (see \cite{lwit}). Therefore, taking the derivative of (\ref{corriente}) with respect to $\tau$, and evaluating it at $\tau=0$ , we have
\begin{equation}
{\cal I} = \frac 1{4\pi } \left( \int_{0}^{2\pi}
\left.{\cal F}^{\dagger}_{02}\right|_{\rho={\cal R}} d\phi+ \int_{S_1} \partial_{t} {\cal F}^{\dagger}_{12} d\rho d\phi  \right)  \; .
\label{corriente2}
\end{equation}

When the fields are stationary the second integral in the right hand side of (\ref{corriente2}) is zero. In this case, since the first integral is taken in the vacuum region, its value is independent of the radius ${\cal R}$, as long as we compute it with the exterior metric. This justifies the procedure given in \cite{lwit}, where the limit $\rho \rightarrow 0$ is given as part of the prescription for what amounts to a definition of ${\cal I}$.  In more detail, consider the stationary exterior solution. Then the non vanishing components of the selfdual electromagnetic tensor are given by

$$
{\cal F}^{\dagger}_{02} = (\Delta - 1) j e^{i \theta _d} \;\;\; , \;\;\; {\cal F}^{\dagger}_{13} = \frac{(\Delta - 1)}{\rho ^ \Delta \Gamma ^2} j e^{i \left( \theta _d + \pi /2 \right)} \; .
$$

Assuming that this tensor is {\em everywhere} stationary, i.e., that $\partial_{t} {\cal F}^{\dagger}_{12}=0$ also in the interior region, the total current may be given as \cite{lwit}
\begin{equation}
{\cal I} = \frac 1{4\pi} \lim_{{\cal R} \rightarrow 0} \int_{0}^{2\pi}
\left.{\cal F}^{\dagger}_{02}\right|_{\rho={\cal R}} d\phi   \; ,
\label{corriente4}
\end{equation}
and we find
\begin{equation}
{\cal I} = \frac{(\Delta - 1)}{2} j e^{i \theta _d}  \; .\label{total}
\end{equation}

We notice that, choosing the rotation angle $\theta_d = 0$ , the resulting  current is purely electric. This is the choice given in, e.g., \cite{lwit,dem}. The same procedure cannot be applied in the general case, where the metric is not stationary, because the first integral in the right hand side of (\ref{corriente2}) is not independent of ${\cal R}$. We might think of using (\ref{corriente4}) to define the current ${\cal I}$, but here the problem is that, as we have shown, for any given set of external parameters, the source region cannot have an arbitrarily small radius if certain physical restrictions hold for the matter and fields inside the source. We may argue, on the other hand, that although we cannot use (\ref{corriente2}) to compute ${\cal I}$, because of the lack of information to compute the integral involving $\partial_{t} {\cal F}^{\dagger}_{12}$, this equation certainly allows for the possibility that ${\cal I}$ is non vanishing. In particular, as we showed above, the metrics we describe in this paper contain as a limit the stationary case, where (\ref{corriente2}) holds. So it is reasonable to look at the right hand side (\ref{corriente4}), for different finite ${\cal R}$, as a measure or indication of the total current flowing in the string, keeping in mind that a definite answer will depend on the detailed model of the source.

An interesting fact that arises when we follow this procedure is that the resulting `total current' is complex, and cannot be made real by a `duality rotation'. At first sight this might be interpreted as an indication that the source must necessarily include `magnetic currents'. We remark, however, that since we are considering only the vacuum region, there is also a different interpretation, that does not require magnetic currents. Restricting for simplicity to Minkowski spacetime, the reasoning is as follows: suppose we have a solution of (\ref{elemag}), with $\rho_m=0$ and non vanishing electric and magnetic currents. Then, we may define a new field $\vec{{\cal B}'}(\vec{x},t)$ by  
$$
 \vec{{\cal B}'}(\vec{x},t) = \vec{{\cal B}}(\vec{x},t)   + \int ^{t}_{t_0}  \vec{j}_m(\vec{x},t') \; dt'   \; ,
 $$
where $t_0$ is in principle arbitrary. The fields $
 \vec{{\cal B}'}(\vec{x},t)$ and $\vec{{\cal B}}(\vec{x},t)$ are then identical outside the spatial support of $j_m$, i.e., for any $\vec{x}_1$ such that $j_m(\vec{x}_1,t)=0$ for all $t$. Then we have
$$
\vec{\nabla} \cdot \vec{{\cal B}'}(\vec{x},t) = \vec{\nabla} \cdot \vec{{\cal B}}(\vec{x},t) +\int ^{t}_{t_0}  \vec{\nabla} \cdot \vec{j}_m(\vec{x},t') \; dt'  \; , 
$$
and, from (\ref{elemag}) and (\ref{conserv}), if $\rho_m=0$ , in terms of $\vec{{\cal E}}$ and $\vec{{\cal B}'}$, we have,
\begin{eqnarray}
\label{elemag1}
\vec{\nabla} \cdot \vec{{\cal B}}' = 0  \nonumber  \;\;\; & , & \;\;\; \vec{\nabla} \cdot \vec{{\cal E}} = 4 \pi \rho _e \; , \nonumber \\
\vec{\nabla} \times \vec{{\cal E}} +  \partial _t \vec{{\cal B}}' = 0  \;\;\; & , &\;\;\; \vec{\nabla} \times \vec{{\cal B}}' - \partial _t \vec{{\cal E}} = 4 \pi \vec{j}'_e  \; , 
\end{eqnarray}
where $\vec{j}'_e$ is an `effective' electric current, given by
$$
\vec{j}'_e = \vec{j}_e + \int ^{t}_{t_0} \vec{\nabla} \times  \vec{j}_m(\vec{x},t') \; dt' \; , 
$$
and we have the conservation equation
$$
\vec{\nabla} \cdot \vec{j}'_e = -  {\partial \rho _e \over \partial t} \; .
$$

Thus, if $\rho_m=0$, the electromagnetic field outside the sources may be considered as part of the solution of either (\ref{elemag}), with both electric and magnetic currents, or of (\ref{elemag1}), with purely electrical currents. Therefore, as remarked above, it is not possible to infer the presence and type of current in the `source' from an analysis of the electromagnetic fields outside this source.

\section{Final Comments}

The exact cylindrically symmetric electrovacuum solution found by L. Witten \cite{lwit} has been considered as representing the electrogravitational field outside a superconducting cosmic string \cite{dem}. This is in agreement with several computations to obtain the metric outside the string \cite{mos,bab}. All these analyses correspond to stationary situations. In this paper we described the construction of solitonic perturbations of the Levi-Civita metric, leading to exact solutions of the Einstein-Maxwell equations, with the appropriate symmetry to be considered as candidates for the metric exterior to a superconducting cosmic string, in the presence of non stationary electromagnetic fields. 

There are a number of issues that arise in trying to make this interpretation concrete. First, we have to deal with the fact that the metrics contain curvature singularities on the symmetry axis, where the 'string' would be located. A simple way of handling this problem is to assume that metric describes the spacetime outside a certain `radius' ${\cal R}_0$. It is then possible to impose some constraints on the parameters by requiring that the `source' (string) for $\rho<{\cal R}_0$ satisfies, e.g, some appropriate energy conditions. In our case this restricts the solutions to the set with $\Delta \geq 1$, in agreement with previous calculations, but furthermore, it also provides a {\em minimal radius} for the source. This leads to the second important issue, namely   that, since the metrics describe non stationary electrovacuum spacetimes, and we need to exclude a tube of finite radius, there is no unique way of computing the current in the source, or even to ascertain if a current is at all present. There is a further complication that stems from the fact that we actually solve Einstein-Maxwell equations for a self-dual field, and, since the potentials include an arbitrary `duality rotation', the `sources' might include magnetic currents. 

The presence of magnetic currents is a well known feature of gauge theories, so this presents no difficulty. However, the point we tried to make is that it is not really possible to decide, just from the external vacuum field, what sort of currents, if any, are present in the source. To make contact with previously accepted superconducting string spacetimes, we noticed that for `wide' solitons, there is range of times where the metric changes slowly with time, and, moreover, it is possible to choose the parameters so that it approaches the stationary solution with non vanishing current.

\section*{Acknowledgments}

This work was supported in part by funds of the University of C\'ordoba, and grants from CONICET and CONICOR (Argentina). R.J.G. is a member of CONICET. M.H.T. holds a doctoral scholarship from CONICOR.

\appendix

\section{Asymptotic behavior at timelike infinity}

At timelike infinity, i.e., fixed $\rho$, $x^1$, and $x^2$, and $t \rightarrow \infty$ , the behavior of the metric depends on the value of $\Delta$. One obtains:

\begin{equation}
\frac{f}{\rho^{(\Delta^2-1)/2}} \approx
\left\{ \begin{array}{lr}
\displaystyle{ \frac{C' \omega ^2 4^{\Delta-1} \rho ^{2-2 \Delta} t^{2 \Delta -4} }{L_0^2} } & \;\; \mbox{if $2<\Delta$} \\
& \\
\displaystyle{ \frac{C'  \left[ 16\omega^2 + \rho ^2  \left(2L_0\sin{\delta_0}-q\right)^2  \right] }{4 \rho^2L_0^2}    } & \;\; \mbox{if $2=\Delta$} \\
& \\
\displaystyle{ \frac{C' (2L_0 \sin{\delta_0} -q)^2}{4L_0^2} } & \;\; \mbox{if $0\leq \Delta<2$} 
\end{array} \right.         \label{latea}
\end{equation}

\begin{equation}
-\frac{g_{22}}{\rho^{1+\Delta}} \approx
\left\{ \begin{array}{lr}
\displaystyle{ \frac{4t^2}{\rho^2}  } & \;\; \mbox{if $2<\Delta$} \\
& \\
\displaystyle{ \frac{64 \omega^2 t^2}{\rho^2 \left[ 16\omega^2 + \rho ^2  \left(2L_0\sin{\delta_0}-q\right)^2  \right] }   } & \;\; \mbox{if $2=\Delta$} \\-
& \\
\displaystyle{ \left(\frac{4t^2}{\rho^2}\right)^{\Delta-1} \frac{16 \omega^2}{\rho^2\left( 2L_0\sin{\delta_0-q} \right)^2}   } & \;\; \mbox{if $1<\Delta<2$} \\  
& \\
\displaystyle{ \frac{  \left[ 16\omega^2 + \rho ^2  \left(2L_0\sin{\delta_0}-q\right)^2  \right] }{\rho^2\left(2L_0\sin{\delta_0}-q\right)^2}  } & \;\; \mbox{if $1=\Delta$} \\
& \\
\displaystyle{ 1  } & \;\;  \mbox{if $0\leq \Delta<1$} 
\end{array} \right.  \label{lateb}
\end{equation}

\begin{equation}
-\frac{g_{33}}{\rho^{1-\Delta}} \approx
\left\{ \begin{array}{lr}
\displaystyle{ \frac{\rho^2}{4t^2} } & \mbox{if $3<\Delta$} \\
& \\
\displaystyle{  \frac{ \rho^2 \left[ 16\omega^2 + \rho ^2 \left(2L_0\sin{\delta_0}-q\right)^2  \right] }{64\omega^2t^2}   }  & \;\; \mbox{if $3=\Delta$} \\
& \\
\displaystyle{ \left(\frac{\rho^2}{4t^2}\right)^{\Delta-2} \frac{ \rho ^2 \left(2L_0\sin{\delta_0}-q\right)^2}{16\omega^2}  }  & \;\; \mbox{if $2<\Delta<3$} \\
& \\
\displaystyle{\frac{\rho^2\left(2L_0\sin{\delta_0}-q\right)^2} {  \left[ 16\omega^2 + \rho ^2 \left(2L_0\sin{\delta_0}-q\right)^2  \right] }   } & \;\; \mbox{if $2=\Delta$} \\
& \\
\displaystyle{ 1  } & \;\; \mbox{if $0\leq \Delta<2$}
\end{array} \right.   \label{latec}
\end{equation}

\begin{equation}
g_{23} \approx \left\{ \begin{array}{lr}
\displaystyle{ \left( \frac{\rho^2}{4t^2} \right)^{(\Delta-3)/2} \frac{\rho^2\left(2L_0\sin{\delta_0}-q\right)}{4\omega}  }  & \mbox{if $2<\Delta$}  \\
& \\
\displaystyle{ \frac{8t \rho \omega \left(2L_0\sin{\delta_0}-q\right)}{ \left[ 16\omega^2 + \rho ^2 \left( 2L_0\sin{\delta_0}-q\right)^2   \right] }  } & \mbox{if $2=\Delta$} \\
& \\
\displaystyle{ \left( \frac{4t^2}{\rho^2} \right)^{(\Delta -1)/2} \frac{4\omega}{2L_0\sin{\delta_0}-q}  } & \mbox{if $0<\Delta<2$} \\ 
& \\
\displaystyle{- \frac{2\rho\omega \left(L_0^2-1\right)}{t\left(2L_0\sin{\delta_0}-q\right)}  } & \mbox{if $0=\Delta$}
\end{array} \right.   \label{lated}
\end{equation}

\section{The nature of the singularities}

We analyze here the cases where $q = 2 L_0 \sin{\delta _0}$ . The condition that $q$ is a non negative constant imposes, in turn, that $L_0 \geq 0$ and $\delta _0 \in [0,\pi]$. Consider first the electrovacuum case, i.e., $L_0 >0$ and $\delta _0 \in (0,\pi)$ . The behavior of the solutions near null or spatial infinity is the same as when $q \neq 2 L_0 \sin{\delta _0}$ . We, therefore, present the behavior at timelike infinity that is obtained imposing from the beginning the condition $q = 2 L_0 \sin{\delta _0}$ (this is necessary, because the limits $t \rightarrow \infty$ and $q \rightarrow  2 L_0 \sin{\delta _0}$ do not commute). The results are:

\begin{equation}
\frac{f}{\rho^{(\Delta^2-1)/2}} \approx
\left\{ \begin{array}{lr}
\displaystyle{\frac{C' D^4 4^\Delta \rho^{2-2 \Delta} t^{2\Delta -4} }{L_0 ^2 \sin{\delta _0} ^2} } & \mbox{if $1<\Delta$} \\
& \\
\displaystyle{\frac{4 C' D^4 \left( 1+2L_0 \sin{\delta _0}+L_0 ^2 \right) } { L_0^2 t^2\sin^2{\delta _0} } } & \mbox{if $1=\Delta$} \\
& \\
\displaystyle{\frac{C' 4 D^4 \left( 1+\Delta ^2 \tan^{-2}{\delta _0} \right)}{t^2} } & \mbox{if $0<\Delta<1$} \\
& \\
\displaystyle{\frac{4 C' D^4} {t^2 } } & \mbox{if $0=\Delta$} \\
\end{array} \right.  \label{Latea}
\end{equation}

\begin{equation}
- \frac{g_{22}}{ \rho ^{1+\Delta}} \approx
\left\{ \begin{array}{lr}
\displaystyle{ \sigma \approx  \frac{4t^2}{\rho ^2} } & \;\;\;\;\;\;\;\;\;\;\;\;\;\;\;\; \mbox{if $0<\Delta$} \\
& \\
\displaystyle{\frac{L_0 ^2 \cos^2 {\delta _0} +1 } {L_0^2 \sin^2 {\delta _0} } } & \;\;\;\;\;\;\;\;\;\;\;\;\;\;\;\; \mbox{if $0=\Delta$} \\
\end{array} \right. \label{Lateb}
\end{equation}

\begin{equation}
- \frac{g_{33}}{ \rho ^{1-\Delta}} \approx 
\left\{ \begin{array}{lr}
\displaystyle{ \frac{\rho ^2}{\sigma} \approx \frac{\rho ^4}{4t^2} } & \mbox{if $2<\Delta$} \\
& \\
\displaystyle{ \frac{\rho ^2 (1+L_0^2 \cos^2{\delta_0}) }{4t^2} } & \mbox{if $2=\Delta$} \\
& \\
\displaystyle{\frac{\rho ^4 (\Delta -1) ^2 } {4t^2 \tan^2 {\delta _0} } } & \mbox{if $1<\Delta<2$} \\
&\\
\displaystyle{ \frac{\rho ^2 (1 + 2L_0 \sin{\delta _0} + L_0^2) }{4 t^2} } & \mbox{if $1=\Delta$} \\
& \\
\displaystyle{ \frac{\rho ^4 (\Delta -1) ^2 } {4t^2 \tan^2 {\delta _0} } } & \mbox{if $0<\Delta <1$} \\
& \\
\displaystyle{ \frac{L_0^2 + \cos^2{\delta _0}}{\sin^2{\delta _0}} } & \mbox{if $0=\Delta$} \\
\end{array} \right.  \label{Latec}
\end{equation}

\begin{equation}
g_{23} \approx 
\left\{ \begin{array}{lr}
\displaystyle{ \frac{ (\Delta-1) 2^{\Delta -2}t^{\Delta -2} \rho ^{3 - \Delta} \cos{\delta _0} }{L_0 \sin{\delta _0}^2} } & \mbox{if $0<\Delta$} \\
& \\
\displaystyle{- \frac{\rho \cos{\delta _0}(1+L_0^2)}{L_0 \sin{\delta _0}^2} } & \mbox{if $0=\Delta$} \\
\end{array} \right.  \label{Lated}
\end{equation}

The behavior on the axis is the same as for $q \neq 2 L_0 \sin{\delta _0}$, except for $\Delta=1$, in which case the analysis of section IV is not valid since the variables given by (\ref{new}) cannot be defined and thus, one cannot fix the topology as in that case. This family of solutions ($\Delta=1$ and $q = 2 L_0 \sin{\delta _0}$) was analyzed in \cite{dgn91}. More details of what follows can be found there. Imposing the condition $q = 2 L_0 \sin{\delta _0}$ right from the beginning, it is found that near the axis

\begin{eqnarray*}
ds^2 & \approx & \frac{C' \omega ^2 (1+2L_0 \sin{\delta _0} + L_0^2 ) }{L_0^2(t^2+\omega ^2)} (-dt^2 +d\rho ^2) \nonumber \\ & & 
- \frac{4(t^2+\omega ^2)}{1+2L_0 \sin{\delta _0} +L_0^2 } dx_2 ^2 - \frac{t^2(1+2L_0 \sin{\delta _0} +L_0^2 ) }{4 (t^2+\omega ^2)} dx_3^2
\end{eqnarray*}
By inspection of the previous equation, one notes that the axis can be made quasiregular (or regular) chosing $x_3 = \phi$ and $x_2=z$, with the usual ranges for these coordinates. Nevertheless, at spatial infinity the roles of the Killing vectors are reversed (see Eqs.(\ref{Lateb}-\ref{Latec})). On the other hand, one can fix the `appropriate' topology at spatial infinity, but then the spacetime does not admit an axis, although it is locally regular. Notice also that, opposite to the case $\Delta =1$ and $q = 2 L_0 \sin{\delta _0}$, here singularities do develop at late times (see Eq.(\ref{lateb})).

Moreover, from (\ref{Lateb}) one notes that when $q = 2L_0 \sin{\delta _0}$ the spacetime becomes singular as $t \rightarrow \infty$ {\em for all} $\Delta>0$, and that the singularity can be reached by observers with constant $\rho$ , $\phi$ and $z$ in finite proper time if $0<\Delta <3/2$ (in the general case, this did not happen for any value of $\Delta$). An idea of what is going on can be obtained analyzing the $\Delta =1$ case, since there a density of energy per unit of length can be defined. We now proceed to do so. 

If  $\Delta=1$, and $q \neq 2L_0 \sin{\delta _0}$, the deficit angle at spatial infinity is $< 2\pi$ and, in particular, it follows that the C-energy is finite. We may consider the induced metric and its extrinsic curvature on a surface of constant $t$, say $t=0$. Then, the whole spacetime can be thought as the evolution of this initial data, and, as we have summarized, no singularities develop. Moreover, it has been shown that under certain general conditions no singularities will develop in electrovacuum gravity from initial data with deficit angle $< 2\pi$ \cite{ber}. When $q \rightarrow 2L_0 \sin{\delta _0}$ this angle approaches $2\pi$ and (thus) the C-energy diverges. In principle, the property that the C-energy diverges is not necessarily pathological. Indeed, this quantity is not the generator of time translations at spatial infinity (except for the weak field limit) and it does not represent the total gravitational mass \cite{ash}. When symmetry reducing a spacetime with translational symmetry, the total Hamiltonian is proportional to the deficit angle at spatial infinity \cite{ash}, which remains finite when this angle approaches the value of $2\pi$. Thus, it cannot be said that when $\Delta =1$ and $q \rightarrow 2L_0 \sin{\delta _0}$ the energy diverges, but rather that it approaches its maximum value. Specifically, in the hamiltonian formulation the constraints for the initial data have solution iff the deficit angle is $< 2\pi$ \cite{ash} and, moreover, whether or not a hamiltonian formulation is imposed, if spatial geodesical completeness is satisfied, then the deficit angle must be $\leq 2\pi$ \cite{carr}. Thus, when $q \rightarrow 2L_0 \sin{\delta _0}$ the total mass approaches its limiting value. 

Similar properties are known for the a ${\cal U}(1)$ stationary gauge string coupled to gravity. When its total mass per unit of length is raised and the string becomes supermassive, there is a curvature singularity at a finite distance of the core (singularities at a finite distance from the core are also present in global strings \cite{glo}). The behavior of the metric near the singularity is as in Levi - Civita spacetime with $\Delta=3$ and $\rho \approx 0$ \cite{sup}. The Levi - Civita with $\Delta =3$ character of the singularity is not so surprising: at large distances from the core the metric must approach a vacuum one with boost invariance along the $z$ direction, and this corresponds to Levi - Civita with either $\Delta=1$ or $\Delta =3$ . The former corresponds to a non supermassive gauge cosmic string, and the later to supermassive ones. The interesting feature is that this singularity is located at a finite distance from the core: the spacetime could have approached a Levi - Civita one with $\Delta =3$ and $\rho \approx \infty$ and it would still be boost invariant but, as was mentioned at the beginning of Section IV, the singularity would not be reachable by test particles or photons in finite affine parameter.

In summary, that the spacetime turns singular at late times for $\Delta =1$ as $q \rightarrow 2L_0 \sin{\delta _0}$ is related to the fact that in this limit the total mass approaches its maximum value. But then it is even more natural that singularities also develop when the metric is singular at spatial infinity, as for $\Delta >1$. In any case, one expects that systems with translational symmetry will approximate others without symmetry only locally and, as mentioned in  Section VI, we are mostly interested on the case $\Delta >1$ and $\Delta \approx 1$, in the region of spacetime near the axis and for not large times.

Finally, note that some of the asymptotic expressions are not valid if $L_0 \sin{\delta _0} =0$. In Appendix C we will carry out these calculations again, imposing from the beginning the condition $L_0=0$. Although the features of these solutions are, basically, the ones that we have found up to this point, we present explicitly these family of solutions because we want to show what we have already remarked in Section IV; in the general case, the behavior near the axis for $\Delta >2$ is the same as for these vacuum families.

\section{The vacuum and diagonal subfamily}

Within the subfamily $q= 2L_0 \sin{\delta  _0}$ we further restrict to $L_0=0$. Then we obtain metrics that constitute the diagonal cases of the vacuum subfamily:

\begin{equation}
ds^2 = \frac{C'}{{\cal H}} \rho^{(\Delta^2-1)/2} \sigma^{\Delta+1}\sin^2{\phi}  (dt^2 - d\rho^2) - \rho^{1+\Delta} \sigma
d\phi^2 - \frac{\rho^{1-\Delta}}{\sigma}dz^2 \; , \label{vacdiag}
\end{equation}
where
$$
{\cal H} =   (\sigma -1)^2 + \frac{16 \sigma ^2 \omega^2}{\rho ^2 (\sigma -1)^2} \; .
$$
We are now able to see, from the previous expressions, that the behavior near the axis for $\Delta >2$ and $q\neq 2L_0 \sin{\delta _0}$ (see Eq.(\ref{axis1})) is exactly the one given by (\ref{vacdiag}). 

Note that it is no longer possible to make the change $L_0 \rightarrow 1/ L_0$ , so the metrics with $\pm \Delta$ are not locally isometric. In fact, we will see that there are differences in the solutions with $\Delta = \pm 1$. 
As in all the previous cases, at spatial infinity the metric approaches the seed {\em for all} $\Delta$, so we can just  insert $q=0=L_0$ in the corresponding expressions of Section IV to obtain the explicit asymptotic behavior in these region.

At timelike infinity the norm of one of the Killing vectors diverges {\em for all} $\Delta$:

\begin{displaymath}
ds^2 \approx C' \omega ^2 4^{\Delta -2} \rho^{(\Delta -1)(\Delta -3)/2} t^{2\Delta -4}  (dt^2 -d\rho ^2) - 4 \rho^{\Delta -1} t^2 dx_2 ^2 - \frac{\rho^{3-\Delta} } {4 t^2} dx_3^2  \; ,
\end{displaymath}
and the metric near the axis is 
\begin{equation}
ds ^2 \approx C' \omega ^2 4^{\Delta -1} \rho ^{(\Delta - 1)(\Delta - 3)/2}  (t^2 + \omega ^2)^{\Delta -2}(dt^2 - d\rho^2) - 4 \rho ^{\Delta -1}(t^2 + \omega^2) dx_2 ^2 - \frac{\rho ^{3 - \Delta} }{4(t^2 + \omega ^2)} dx_3 ^2  \; ,      \label{eje2}
\end{equation}
and it can be chosen regular or quasiregular iff $\Delta =1,3$ . We do not repeat the analysis because it is the same as when $q=2L_0 \sin{\delta  _0}$ and $L_0 \neq 0$, i.e.,  near the axis the solution is locally regular, but at spacelike infinity the roles of the Killing vectors are reversed. On the other hand, when $\Delta=-1$, the axis is singular, and it can be explicitly seen that there are curvature scalars that diverge as this axis is approached \cite{dgn91}, although at spatial infinity the metric becomes flat.

With regard to the singularities for $t \rightarrow \infty$ , when $\Delta \geq 1$, observers with constant $\rho$, $\phi$ and $z$ have infinite proper time when $t \rightarrow \infty$, but there are timelike curves that reach the singularity with finite proper time.


\begin{thebibliography}{99}

\bibitem{ewit} E. Witten, Nucl. Phys. {\bf B 249}, 557 (1985)

\bibitem{car} B. Carter, Int. J. Theor. Phys. {\bf 36}, 2451 (1997).

\bibitem{lwit} L. Witten, in {\em Gravitation: an Introduction to Current Research} (John Wiley \& Sons, New York, 1962).

\bibitem{par} C. T. Hill, H. M. Hodges, and M. S Turner, Phys. Rev. D {\bf 37}, 263 (1988); P. Amsterdamski and P. Laguna-Castillo, Phys. Rev. D {\bf 37}, 877 (1988); A. Babul, T. Piran and D. N. Spergel, Phys. Lett {\bf B202}, 307 (1988).

\bibitem{pet} P. Peter, Phys. Rev. D {\bf 49}, 5052 (1994).

\bibitem{lev}  T. Levi-Civita, Rend. Acc. Lincei, {\bf 28}, 3 (1919)

\bibitem{ale}  G. A. Alekseev, Sov. Phys. JETP {\bf 32} (1980); Proc. Steklov Inst. Math {\bf 3}, 215 (1987).

\bibitem{ver} E. Verdaguer, Phys. Rep. {\bf 229} (1993).

\bibitem{ver2}  B. J. Carr and E. Verdaguer, Phys. Rev. D {\bf 28}, 2995 (1983).

\bibitem{dgn90} A. D. Dagotto, R. J. Gleiser, and C. O. Nicasio, Phys. Rev D {\bf 42}, 424 (1990).

\bibitem{cil} J. Colding, N. K. Nielsen, and Y. Verbin, Phys. Rev. D. {\bf 56}, 3371 (1997); V. P. Unruh, W. Israel, and W. G. Unruh, Phys. Rev. D. {\bf 39}, 1084 (1989).

\bibitem{mos} I. Moss and S. Poletti, Phys. Lett. B {\bf 199}, 34 (1987).

\bibitem{tho} K. S. Thorne, Phys. Rev. {\bf 138}, 251 (1965).

\bibitem{dem}  M. Demianski, Phys. Rev. D {\bf 38}, 698 (1988).

\bibitem{aut} W. B. Bonner, Proc. Phys. Soc. London {\bf 67A}, 225 (1953); M. Misra and L. Radhakrishna, Proc. Natl. Inst. Sci. India {\bf 28A}, 632 (1962); B. K. Harrison Phys. Rev. D. {\bf 138}, B488 (1965).

\bibitem{mel} M. A. Melvin, Phys. Lett. {\bf 8}, 65 (1964).

\bibitem{gar}  D. Garfinkle, Phys. Rev. D {\bf 32}, 1323 (1985).

\bibitem{bab} A. Babul, T. Pirani, and D. N. Spergel, Phys. Lett. B {\bf 209}, 477 (1988).

\bibitem{dgn91} A. D. Dagotto, R. J. Gleiser, and C. O. Nicasio, Phys. Rev D {\bf 43}, 1162 (1991).

\bibitem{ber} B. K. Berger, P. T. Chru\'sciel, and V. Moncrief, Ann. Phys {\bf 237}, 322 (1995).

\bibitem{ash} A. Ashtekar and M. Varadarajan, Phys. Rev. D {\bf 50}, 4944 (1994).

\bibitem{carr} S. M. Carroll, E. Farhi, A. H. Guth, and K. D. Olum, Phys. Rev. D {\bf 50}, 6190 (1994). 

\bibitem{glo} G. W. Gibbons, M. E. Ortiz, and F. R. Ruiz, Phys. Rev. D {\bf 39}, 1546 (1989); D. Harari and P. Sikivie, Phys. Rev. D {\bf 37}, 3438 (1988); R. Gregory, Phys. Lett {\bf B 215}, 663 (1988); A. G. Cohen and D. B. Kaplan, Phys. Lett {\bf B 215}, 67 (1988).  

\bibitem{sup} P. Laguna and D. Garfinkle, Phys. Rev. D {\bf 40}, 1011 (1989).

\end{thebibliography}
\end{document}